\documentclass[a4paper,amsmath,preprintnumbers,showpacs,twocolumn,prl, nofootinbib]{revtex4}
\usepackage{}
\usepackage{amssymb}
\usepackage[hypertex]{hyperref}
\usepackage{graphicx}
\usepackage{epstopdf}
\usepackage{bm}
\usepackage{epsfig}
\usepackage{graphics}
\usepackage{xspace}
\usepackage{amsfonts}
\usepackage{verbatim}
\usepackage{natbib}
\usepackage{color}

\graphicspath{{./eps/}}

\newcommand{\nn}{\nonumber}

\begin{document}


\preprint{\font\fortssbx=cmssbx10 scaled \magstep2
\hbox to \hsize{
\hfill$\vcenter{\hbox{ MITP/13-069}
                }$}
}

\title{Determining the masses of invisible particles: Application to Higgs boson invisible decay}

\author{Jian Wang}\email{jian.wang@uni-mainz.de}
\affiliation{PRISMA Cluster of Excellence $\&$ Mainz Institute for Theoretical Physics,
Johannes Gutenberg University, D-55099 Mainz, Germany }

\date{\today}



\begin{abstract}

To know the total width of the recently discovered Higgs boson particle, it is important to measure the invisible decay width of the Higgs boson.
However, the signal for this measurement at the LHC,
i.e., a charged lepton pair and missing energy in the final state, cannot be definitely understood as the product of
the intermediate produced $Z$ and $H$ bosons due to the possible interaction between dark matter and a $Z$ boson or quarks,
which can be described by representative effective operators.
First, we consider the relic abundance, the LUX direct detection experiment and the result of searching for Higgs
boson invisible decay at the LEP and LHC to find the allowed parameter region for the effective operators.
Then we investigate the transverse momentum distribution of the missing energy and propose two observables that can be used to distinguish
the different underlying processes.
Moreover, with these two observables, we may be able to determine the masses of invisible particles.

\end{abstract}

\maketitle


\section{Introduction}

It is important to precisely measure the properties of the Higgs boson after its discovery \cite{Aad:2012tfa,Chatrchyan:2012ufa}.
The total width of the Higgs boson is difficult to measure at a hadron collider due to the unmeasurable partonic center-of-mass energy.
As a result, one can only get information on the total width from the global fit and sum over various decay channels of the Higgs boson \cite{Barger:2012hv}
\footnote{We notice that a constraint is presented on the total width of the Higgs boson,
i.e., $\Gamma_{\rm H}< 22 {\rm ~MeV}$ at the $95\%$ confidence level \cite{Khachatryan:2014iha},
by using its relative on-shell and off-shell production and decay rates to a pair of $Z$ bosons \cite{Caola:2013yja}. }.
The largest decay channel is $H\to b\bar{b}$, which suffers from overwhelming QCD backgrounds.
Although it becomes possible to detect $H\to b\bar{b}$ by using a delicate method proposed in Ref. \cite{Butterworth:2008iy}
and the ATLAS and CMS collaborations have searched for
the $b\bar{b}$ decay in the $V(W,Z)H$ associated productions, the statistical uncertainties of the results are still too large \cite{ATLAS-CONF-2013-079,CMS-PAS-HIG-13-012}.
The decay channel $H\to gg$ faces the same problem, but there is no b-tagging technology that can be used in $H\to b\bar{b}$.
The decay width of this channel can only be obtained from a global fit of the relevant Higgs couplings.

In addition, there are still possibilities for the Higgs boson decaying to invisible particles \cite{Shrock:1982kd,Choudhury:1993hv,Gunion:1993jf,Eboli:2000ze,Belotsky:2002ym,Godbole:2003it,Davoudiasl:2004aj,Patt:2006fw,Aad:2012tfa,Chatrchyan:2012ufa}.
To measure the invisible decay width of the Higgs boson at colliders, it is necessary to consider the associated production of the Higgs boson
with some visible particles. The $Z$ boson is a good choice due to its large coupling to the Higgs boson,
large production rate and clear signature at colliders.
Because of the limited center-of-mass energy, the experiment at the LEP has only excluded the Higgs boson mass range below 114.4 GeV
via the Higgs-strahlung process $e^+e^- \to HZ$ \cite{Searches:2001ab}.
Recently, the ATLAS collaboration at the LHC carried out a similar search in the $HZ$ associated production with $H$ decaying invisibly and the $Z$
boson decaying into a charged lepton pair.
The present result shows no deviation from the Standard Model (SM) expectation and
constrains the invisible branching fractions to be less than $65\%$ at the $95\%$ confidence level \cite{ATLAS-CONF-2013-011}.
The CMS collaboration has obtained a similar result \cite{CMS-PAS-HIG-13-018}.

We are interested in the question of, if some deviation from the SM expectation is observed in the future as the integrated luminosity of the LHC
is increased, if can we determine the invisible decay width of the Higgs boson.
In fact, the experimentally observed final state is just a charged lepton pair and missing energy.
And the analysis is based on the assumption that a resonance and a $Z$ boson have been associated produced with the cross section predicted by the SM
and the resonance totally decaying invisibly.
It is reasonable that the charged lepton pair can be attributed to the decay product of the intermediate $Z$ boson
if the invariant mass of the charged lepton pair is around the $Z$ boson mass $M_Z$.
However, it is not very convincible to interpret the missing energy as a decay product of the intermediate Higgs boson
because there are other possible origins of the missing energy.
For example, it is likely that some dark matter (DM) can interact  with the $Z$ boson or quarks,
which has been also studied extensively in the recent years.
They will appear at the LHC with the same signature.
Thus, it is essential to extract more information about the missing energy.

In this work, we perform such a study toward this direction, focusing on the mass of the missing particles.
If one can determine the mass of the missing particle,
one can be more confident to judge whether the intermediate particle is the Higgs boson or not.
The candidate for the Higgs boson decay products should have a mass lower than one-half of the Higgs boson mass.
This study is closely related to searching for new physics.
In the $R$-parity violating supersymmetry model,
some sparticles would decay into SM particles and a neutralino, which becomes invisible if it is lighter than the other supersymmetry particles
and stable enough \cite{Barbier:2004ez}.
In the large extra dimension model, the Kaluza-Klein graviton can escape from the detection at colliders, manifesting itself as missing energy \cite{Giudice:1998ck}.
Cosmology observation has confirmed the existence of DM in our Universe \cite{Hinshaw:2012aka}.
It can be produced at colliders if it is light and has interactions with SM particles.
Because it is stable, it is not detectable at colliders \cite{Cao:2009uw}.
Searching for all these kinds of new physics requires analyzing  the events with missing energy and recoiling particles,
e.g., a photon \cite{Gao:2009pn,Wang:2012qi,Huang:2012hs,Chatrchyan:2012tea,Aad:2012fw},
a lepton \cite{Bai:2012xg},
a jet \cite{Karg:2009xk,Fox:2012ru},
a $W/Z$ boson \cite{Kumar:2010ca,Bell:2012rg,Carpenter:2012rg},
or a top quark \cite{Andrea:2011ws,Wang:2011uxa}.
Therefore, it would be helpful to distinguish them if we can know more about the missing energy.
In this paper, we focus on the case of missing energy and a $Z$ boson associated production,
which is important to determine the invisible decay width of the Higgs boson.
The method employed here can be generalized to other cases.

\begin{figure}
  \includegraphics[width=0.95\linewidth]{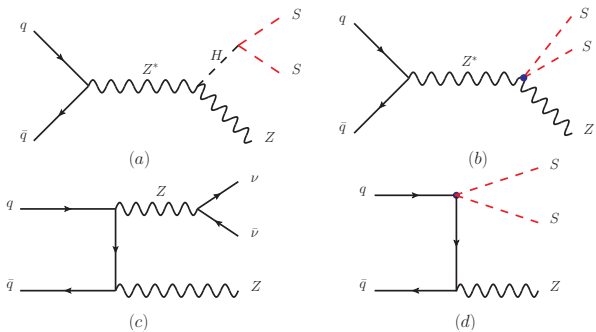}\\
  \caption{Leading Feynman diagrams for a $Z$ boson and missing energy associated production.
  The diagram $(a)$ is the signal process in the analysis of the ATLAS experiment and
  the diagram $(c)$ is the main irreducible background.
  The diagrams $(b)$  and $(d)$ are induced by the effective operators in Eq.(\ref{eqs:oz}).  }
  \label{fig:ZH}
\end{figure}

\section{Effective operators}

The Higgs boson is special in the SM.
Its mass term is not fixed by the gauge invariance and provides an opportunity for this particle to couple with some
SM gauge singlets still with renormalizable interactions \cite{Patt:2006fw}.
The simplest case is adding to the SM a new real scalar $S$ with the Lagrangian given by
\cite{Silveira:1985rk,McDonald:1993ex,Burgess:2000yq,Barger:2007im}
\begin{equation}
    \mathcal{L}_{\rm min}=\mathcal{L}_{\rm SM}+\frac{(\partial_{\mu} S)^2}{2}
    -\frac{m_0^2}{2}S^2 - \lambda S^2 |H|^2 - \frac{\lambda_S}{4!}S^4.
\end{equation}
The new real scalar can account for the observed DM density if a global $Z_2$ symmetry is imposed,
under which all the SM particles are singlets while the new scalar can transform nontrivially, i.e.,$S\to -S$.
To maintain this $Z_2$ symmetry, the new scalar should have no vacuum expectation value (VEV), $\langle S \rangle=0$.
Thus, its mass $m_S$ is given by $m_S^2=m_0^2+\lambda v^2$, where $(0,v)^T/\sqrt{2}$ is the VEV of the Higgs field.
When $2m_S$ is smaller than the Higgs boson mass $m_H$, the Higgs boson can decay into an $S$ pair, namely, Higgs boson invisible decay.
This would modify the total width of Higgs boson and therefore affect the cross section of Higgs boson production and decay into other final states.
Through the mixing with the Higgs boson, the $S$ pair can annihilate to SM particle pairs,
for which the cross section is constrained by the DM relic density.
With the same interaction as in annihilation, $S$ can scatter with nucleons, which can be measured by DM direct detection experiments.
Given that the Higgs boson mass is around 125 GeV, this model receives strong constraint
after considering the results of searching for Higgs boson at the LHC, cosmological relic density and
the DM direct detection \cite{Raidal:2011xk,Mambrini:2011ik,He:2011de,Fox:2011pm,Low:2011kp,Djouadi:2011aa,Cheung:2012xb}.
Therefore, more complicated models ,
in which more particles are included, are proposed \cite{Abada:2011qb,Pospelov:2011yp,Abada:2013pca,Greljo:2013wja}.
There are no rules, in principle, that these additional particles should be very light \cite{Abada:2011qb}.
And they are perhaps heavy since a lighter particle is often much easier to find either in the decay product or direct production at colliders.
In this case, the role played by these particles can be described by effective operators,
\begin{eqnarray}
\label{eqs:oz}
  & & \mathcal{O}_Z=\frac{m_Z^2}{4\Lambda_Z^2} Z^{\mu}Z_{\mu} S^2  ,   \qquad   \mathcal{O}_q=\frac{m_q}{2\Lambda_q^2}\bar{q} q S^2.
\end{eqnarray}
They can induce the production processes described by the Feynman diagrams $(b)$ and $(d)$ in Fig.\ref{fig:ZH},
generating the same signature at hadron colliders as the process assumed in the experimental analysis, e.g., the Feynman diagrams $(a)$ and $(c)$.
Note that $\mathcal{O}_Z$ and $\mathcal{O}_q$  are not gauge invariant.
In fact, one can write down gauge invariant operators with higher dimensions and get the above two operators after symmetry breaking.
For example, the first operator $\mathcal{O}_Z$ can be generated from the gauge invariant dimension-6 effective operator:
\begin{equation}\label{eqs:OZdim6}
    \mathcal{O}_Z^{(6)}= \frac{\kappa}{\Lambda^2} (D^{\mu}H)^{\dagger}(D_{\mu}H)S^2,
\end{equation}
where $D_{\mu}$ is the usual covariant derivative, and $H$ denotes the SM Higgs field.
After the Higgs field gets a nonvanishing vacuum expected value,
$\mathcal{O}_H^{(6)}$ would deduce to $\mathcal{O}_Z$ and the associate similar operator
\begin{equation}
\mathcal{O}_W=\frac{M_W^2}{2\Lambda_W^2} W^{+\mu}W^{-}_{\mu} S^2
\end{equation}
because the Higgs field couples with $W$ and $Z$ bosons simultaneously.
The second operator $\mathcal{O}_q$ can be generated from the gauge invariant operator of dimension 6
\begin{equation}
    \mathcal{O}_q^{(6)} = \frac{\kappa \lambda_q }{\Lambda^2} \bar{Q}_LH q_R S^2 + H.c.,
\end{equation}
where $\lambda_q$ is the Higgs-quark-quark Yukawa couplings in the SM.

Here we choose these two effective operators in Eq.(\ref{eqs:oz})
because they are the only ones in the leading power order, and representative,  one of them inducing an $s$-channel and
the other inducing a $t$-channel process; see Feynman diagrams $(b)$ and $(d)$ in Fig.\ref{fig:ZH}.
The detailed discussion on the possible gauge symmetry breaking mechanism is beyond the scope of this paper.
The above effective operators  have been discussed in the study of DM \cite{DelNobile:2011uf,Goodman:2010ku}.

The main difference between Feynman diagrams $(a)$ and $(b)$ is whether the missing particles have a fixed invariant mass.
In some phase space points where the invariant mass of the $S$ pair is around the Higgs boson mass,
 Feynman diagram $(b)$ would be equivalent to Feynman diagram $(a)$ up to some constant coefficients.
However, the phase space for missing particles cannot be fixed experimentally and must be integrated in any observable.
It is the purpose of this study to find a way to reveal this difference.

\section{Numerical Discussion}

To distinguish the different processes shown in Fig.\ref{fig:ZH} and extract precise information on Higgs boson invisible decay,
we should understand the properties of the cross sections first.
At the moment, we neglect the interference among the Feynman diagrams $(a,b,c,d)$.
This is reasonable because the total widths of the Higgs and $Z$ bosons are small enough
so that we can use narrow width approximation in calculating the Feynman diagrams $(a,c)$
and thus do not need to consider their interference with the Feynman diagrams $(b,d)$.
In numerical discussion, we have taken $m_H=125$ GeV, $M_Z=91.18$ GeV.
We use the Monte Carlo method to calculate the cross section for $Z$ boson and missing energy associated production at the LEP and LHC, and find agreement with MadGraph5v1.3.3 \cite{Alwall:2011uj} for processes at the LHC.
The CTEQ6L1 parton distribution function (PDF) set \cite{Pumplin:2002vw} is chosen and
the renormalization and factorization scales are set to be 200 GeV,
which is about the sum of $m_H$ and $M_Z$.

Before proceeding, we have to find the allowed parameter space of ($\Lambda_{Z/q}, m_S$).
In the following, we will discuss the DM relic abundance and direct detection experiment LUX
as well as the $Z$ boson and missing energy associated production at the LEP and LHC.

\subsection{Constraint from relic abundance}
The DM relic abundance is a precision observable in
cosmology and imposes constraint on any DM model.
It is determined by the annihilation cross section of DM to SM particles,
 given by \cite{Kolb:1990vq}
\begin{equation}
    \Omega_{\rm DM} h^2 \approx \frac{1.07\times 10^9 {\rm GeV}^{-1}x_f}{M_{\rm Pl}g_{*}^{1/2}(a+3b/x_f)},
\end{equation}
where $\Omega_{\rm DM}$ is the cold DM energy density of the Universe
normalized by the critical density, and $h=0.700 \pm 0.022$
is the scaled Hubble parameter.
 $a$ and $b$ are the coefficients in the partial wave expansion of the DM annihilation cross section,
$\sigma_{\rm an}v_{\rm M\o l}=a+bv_{\rm M\o l}^2+O(v_{\rm M\o l}^4)$.
$v_{\rm M\o l}$ is called M$\o$ller velocity, defined as \cite{Gondolo:1990dk}
\begin{equation}
    v_{\rm M\o l}=\sqrt{|\textbf{v}_1-\textbf{v}_2|^2-|\textbf{v}_1 \times \textbf{v}_2|^2},
\end{equation}
where $\textbf{v}_1$ and  $\textbf{v}_2$ are the velocities of colliding DMs in the cosmic comoving frame.
Note that this velocity is different from that in collisions at colliders, since
the colliding DMs in the Universe are not necessarily moving along a line.
And $v_{\rm M\o l}$ cannot be transformed into the center-of-mass frame of the two colliding DMs because
the thermally averaged total annihilation cross section $\langle \sigma_{\rm an}v_{\rm M\o l} \rangle$
has to be evaluated in a common frame for all collisions.
In general, it is difficult to calculate $\langle \sigma_{\rm an}v_{\rm M\o l} \rangle$ due to the complex definition of $v_{\rm M\o l}$.
Fortunately, it is proved that \cite{Gondolo:1990dk}
\begin{equation}
    \langle \sigma v_{\rm M\o l} \rangle = \langle \sigma v_{\rm lab} \rangle^{\rm lab}\neq \langle \sigma v_{\rm cm} \rangle^{\rm cm},
\end{equation}
in which $v_{\rm lab}=|\textbf{v}_{1,\rm lab}-\textbf{v}_{2,\rm lab}|$ is
the relative velocity in the rest frame of one of the incoming particles
and $v_{\rm cm}$ is the velocity in the center-of-mass frame of the two colliding DMs.
The DM is moving at nonrelativistic velocities when freezing out; thus, $v\ll 1$.
$g_*$ is the number of relativistic degrees of freedom available at the freeze-out epoch $x_f$.
And $x_f$ is evaluated by \cite{Kolb:1990vq}
\begin{eqnarray}
  x_f &=& \ln A - 0.5 \ln \ln A +  \ln \left( 1+\frac{6b}{a\ln A} \right)
\end{eqnarray}
with $A=0.038ag/g_*^{1/2}M_{\rm Pl}m_S$. $g$ counts the internal degree of freedom and is equal to 1 for a real scalar in our case.

For $SS\to  ZZ$ or $q\bar{q}$, the annihilation cross section is
\begin{eqnarray}
    \sigma_{\rm an}^Z&=&\frac{1}{32\pi s}\left( \frac{M_Z^2}{\Lambda_Z^2} \right)^2\frac{\sqrt{s-4M_Z^2}}{\sqrt{s-4m_S^2}}
    \left( \frac{s^2}{4M_Z^4}-\frac{s}{M_Z^2}+3 \right), \nn\\
    \sigma_{\rm an}^q&=&\frac{N_c}{8\pi}\left( \frac{m_q}{\Lambda_q^2} \right)^2\frac{\sqrt{s-4m_q^2}}{\sqrt{s-4m_S^2}} \frac{s-4m_q^2}{s}.
\end{eqnarray}
These results are in agreement with those in Refs.\cite{Beltran:2008xg,Cao:2009uw,Yu:2011by}.
Substituting $s\approx 4m_S^2 + m_S^2 v^2 + 3 m_S^2 v^4/4$, we obtain
\begin{eqnarray}
  a^Z &=&\frac{1}{32\pi }\left( \frac{M_Z^2}{\Lambda_Z^2} \right)^2 \frac{\sqrt{m_S^2-M_Z^2} \left(4 m_S^4-4m_S^2 M_Z^2+3
   M_Z^4\right)}{2m_S^3 M_Z^4} , \nn \\
  b^Z &=&\frac{1}{32\pi }\left( \frac{M_Z^2}{\Lambda_Z^2} \right)^2\frac{ 3(4 m_S^4-8 m_S^2 M_Z^2+5 M_Z^4)}{16 m_S^3 M_Z^2 \sqrt{m_S^2-M_Z^2}}. \nn\\
\end{eqnarray}
and
\begin{eqnarray}
  a^q &=&\sum_{q}^{m_q<m_S} \frac{N_c}{4\pi}\left( \frac{m_q}{\Lambda_q^2} \right)^2 \left( 1-\frac{m_q^2}{m_S^2} \right)^{3/2} , \nn \\
  b^q &=&\sum_{q}^{m_q<m_S} \frac{N_c}{4\pi}\left( \frac{m_q}{\Lambda_q^2} \right)^2 \left( 1-\frac{m_q^2}{m_S^2} \right)^{1/2}
  \frac{5m_q^2/m_S^2-2}{8}.\nn\\
\end{eqnarray}

\begin{figure}
  \includegraphics[width=0.89\linewidth]{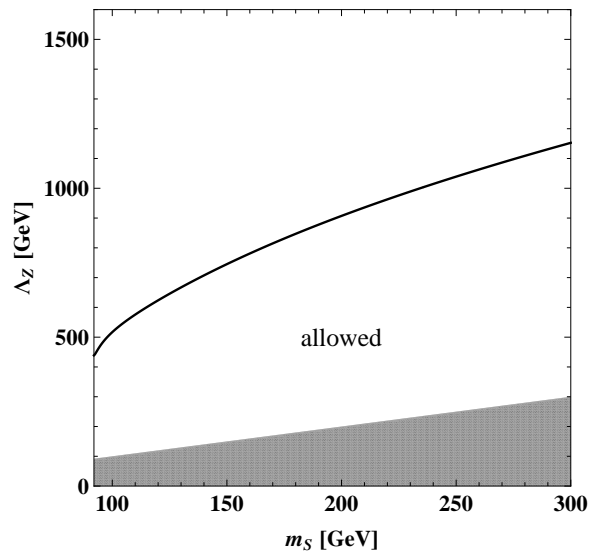}\\
  \caption{The allowed space of ($\Lambda_{Z}, ~m_S$) from relic abundance .
  The region below the solid blue line is allowed. The gray region corresponds to $\Lambda_Z < m_S$.
   }
  \label{fig:relicZ}
\end{figure}

\begin{figure}
  \includegraphics[width=0.89\linewidth]{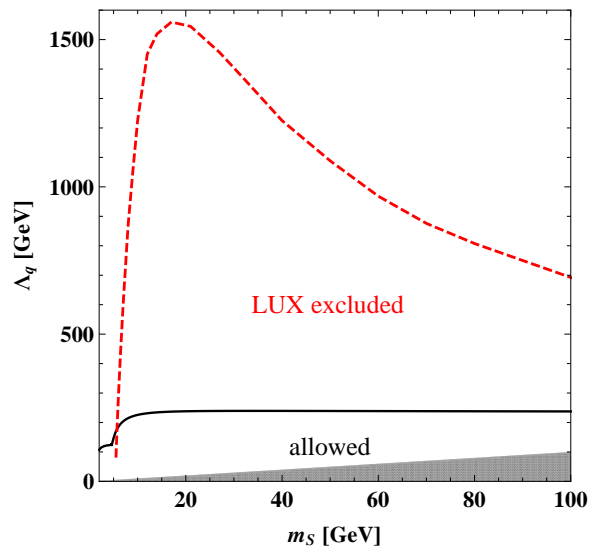}\\
  \caption{The allowed space of ($\Lambda_{q}, ~m_S$) from relic abundance and the direct detection experiment.
  The region below the solid blue line is allowed by the relic abundance.
  The region below the dashed red line is excluded by the LUX direct detection experiment \cite{Akerib:2013tjd}.
  The gray region corresponds to $\Lambda_q < m_S$.
 }
  \label{fig:relic}
\end{figure}

The latest WMAP data and distance measurements from baryon acoustic oscillations in the distribution of
galaxies and Hubble constant measurements give the constraint \cite{Hinshaw:2012aka}
\begin{equation}
    \Omega_{\rm DM} h^2 = 0.1157 \pm 0.0023.
\end{equation}
Requiring that the DM relic abundance is lower than the observed central value at the $2\sigma$ level,
the allowed parameter space of the effective operator is determined.
We show the  parameter space in Figs. \ref{fig:relicZ} and \ref{fig:relic} for the operators $\mathcal{O}_Z$ and $\mathcal{O}_q$
, respectively.
Because of the heavy mass of the $Z$ boson, we show in Fig. \ref{fig:relicZ} only the DM mass larger than 92 GeV.
For $m_S=100$ GeV, the upper limit of $\Lambda_Z$ is about 500 GeV.
For $m_S < M_Z $, the annihilation cross section of $SS\to ZZ$ is vanishing.
Another annihilation channel is required to satisfy the relic abundance constraint, e.g. $SS \to e^+ e^-$.
This can be established in a UV-complete model.
We are working in the effective operator picture here and will not discuss this further.
What we must keep in mind is that the relic density provides an upper limit on  $\Lambda_Z$
while the direct detection experiments and processes at colliders impose lower limits.
There is some parameter space for the DM signal mimicking the Higgs invisible decay
as long as the upper limit is not smaller than the lower limit.

\subsection{Constraint from direct detection}
DM around the Earth can scatter elastically with atomic nuclei, resulting in recoiling movements of nuclei.
These events, if observed, can be explained by DM-nucleon collisions, which can
be divided into spin dependent and spin independent according to the DM-quark interactions.
Generally, the spin independent elastic scattering cross section has a large value and thus gets a more stringent constraint from direct detection experiments.
The most stringent limit on the spin independent elastic scattering cross section comes from the LUX experiment \cite{Akerib:2013tjd}.
Therefore in this analysis, we only consider the data of LUX in order to obtain the allowed parameter space.

The DM-proton spin independent elastic scattering cross section
 \footnote{In practice, the DM-neutron spin independent elastic scattering cross section is almost identical to the DM-proton one \cite{Cao:2009uw}. }
 is given by \cite{Belanger:2008sj,Cao:2009uw}
\begin{equation}
    \sigma_{Sp}^{SI}=\frac{m_p^2}{4\pi(m_S+m_p)^2}[f_{Sp}^{(p)}]^2,
\end{equation}
where
\begin{equation}
    f_{Sp}^{(p)}=\sum_{q=u,d,s} f_{T_q}^{(p)}C_{Sq}\frac{m_p}{m_q}
    +\frac{2}{27}f_{T_g}^{(p)}\sum_{q=c,b,t}C_{Sq}\frac{m_p}{m_q},
\end{equation}
with \cite{Ellis:2000ds}
\begin{eqnarray}
   && f_{T_u}^{(p)}\approx 0.020\pm 0.004, \quad
    f_{T_d}^{(p)}\approx 0.026\pm 0.005, \quad   \nn\\
   && f_{T_s}^{(p)}\approx 0.118\pm 0.062, \quad
      f_{T_g}^{(p)}\approx 1- f_{T_u}^{(p)}- f_{T_d}^{(p)}- f_{T_s}^{(p)}. \nn\\
\end{eqnarray}
In our case, $C_{Sq}=m_q/\Lambda_q^2$.
After comparing with the LUX data \cite{Akerib:2013tjd}, we obtain the allowed parameter region in Fig. \ref{fig:relic}.
It is observed that the combination of the relic abundance and direct detection experiment
results in very stringent constraints on $\mathcal{O}_q$.
Only a limited parameter space with $m_S<5.5$ GeV and $\Lambda_q<$110 GeV is allowed.
We notice that there are no such constraints on the parameters of the operator $\mathcal{O}_Z$
since a nucleon does not contain a $Z$ boson.
Therefore, we will discuss only the case of $\mathcal{O}_Z$ in the following part.

\subsection{Constraints from $Z$ boson and missing energy associated production}
\begin{figure}
   \includegraphics[width=0.89\linewidth]{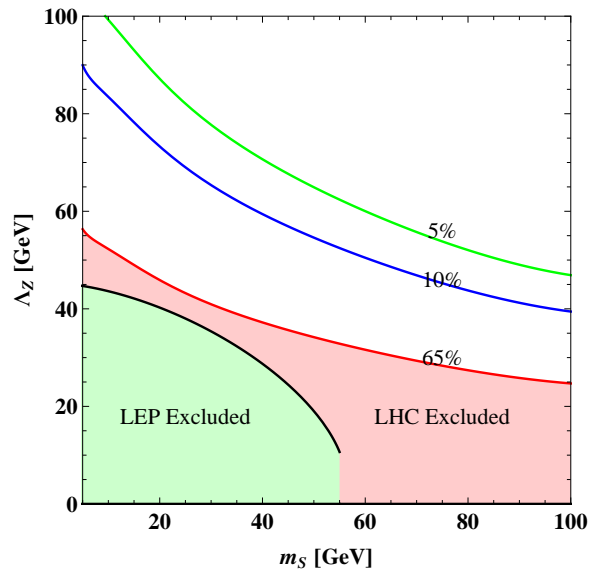}\\
  \caption{The allowed space of ($\Lambda_{Z}, ~m_S$) from $Z$ boson and missing energy associated production.
  The shaded region is excluded by the limit on the missing energy and $Z$ boson production at the
   LEP \cite{Searches:2001ab} and the LHC \cite{ATLAS-CONF-2013-011}.
  The three curves from the bottom up  correspond to the cross section of $HZ$ production at the LHC
  with a Higgs boson branching fraction of $65\%,10\%,{\rm ~and ~}5\%$, respectively.}
  \label{fig:totX}
\end{figure}

Apart from DM annihilation and elastic scattering with nucleons, DM can be produced at the colliders.
Here we are interested in the process of $Z$ boson and missing energy associated production at the LEP and LHC,
i.e., $e^+ e^- (pp) \to Z^* \to SSZ$.
For the LEP experiments, we take the upper limit on the invisible Higgs boson production cross section at 206.0 GeV,
given in Ref. \cite{Searches:2001ab}.
We show the excluded parameter space in Fig. \ref{fig:totX}.
It is found that the region excluded by the current LHC has covered all that by LEP,
indicating the better sensitivity of the LHC to the new physics.
As the increasing of the data accumulated at the LHC,
the excluded region will enlarge if no signal of new physics is observed.
We show in Fig. \ref{fig:totX} as well the curves corresponding to the upper limit on the cross section  of $HZ$ production at the LHC
with a Higgs boson branching fraction of $65\%,10\%, {\rm ~and~} 5\%$, respectively.

On the other hand, if an excess of events is found in the future,
it may be induced by $\mathcal{O}_Z$ with the parameter combination lying on the curve
that corresponds to a fixed Higgs boson branching fraction.
Then it is essential to judge that this excess results from Higgs boson invisible decay or DM associated production with a $Z$ boson.

\subsection{Missing transverse momentum distribution}

\begin{figure}
\includegraphics[width=0.99\linewidth]{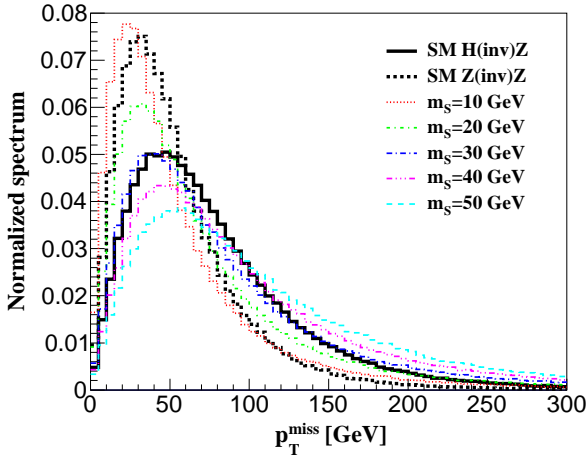}\\
  \caption{The $p_T^{\rm miss}$ distribution for missing energy and $Z$ boson  production at the 8 TeV LHC induced by $\mathcal{O}_Z$.
  We also show the SM processes of $H({\rm invisible~decay})Z$ and $Z({\rm invisible~decay})Z$  production. }
  \label{fig:pt_Z_SSZB}
\end{figure}

Now we show the normalized $p_T^{\rm miss}$ distributions for  missing energy and $Z$ boson  production    at the 8 TeV LHC
in Fig.\ref{fig:pt_Z_SSZB} for the process induced by $\mathcal{O}_Z$.
We see that the shape of the process induced by $\mathcal{O}_Z$
with $m_S=$ 30 GeV is very similar to the SM process of $H({\rm invisible~decay})Z$ production.
This similarity suggests that one cannot simply interpret the signal of missing energy and a charged pair production
as the associated $HZ$ production with Higgs boson invisible decay.

From Fig.\ref{fig:pt_Z_SSZB}, we find that the peak position for the process induced
by $\mathcal{O}_Z$ is moving toward the  large $p_T^{\rm miss}$ region,
and the tail of the distribution drops more slowly as the DM mass increases.
To make it explicit, we present the fitted formulas describing the peak positions:
\begin{eqnarray}
\label{eqs:pTpeak1}
 p_{T,\mathcal{O}_Z}^{\rm peak} & = & 15.83 {~\rm GeV} + 6.67{~\rm GeV} \frac{m_S}{10\rm~GeV}.
\end{eqnarray}
We notice that this formula is fitted from the partonic simulation results, not including the parton shower and hadronization effects,
high-order QCD corrections, etc.
The exact peak positions may be different after taking into account all these effects.
However, we still use this formula as long as the difference between the fitted value and the exact one is not very large.
For comparisons, we also list the values of the peak positions for the processes in the SM:
\begin{eqnarray}
\label{eqs:pTpeak2}
 p_{T,HZ}^{\rm peak} & = & 50.0 {~\rm GeV},  \\
 p_{T,ZZ}^{\rm peak} & = & 32.5 {~\rm GeV}.
 \label{eqs:pTpeak3}
\end{eqnarray}

At the low integrated luminosity of the LHC, the observed events may not be enough to provide a full description of the $p_T^{\rm miss}$ distribution.
For this reason, we propose two observables that can be used to distinguish the underlying processes at the early stage of the LHC, defined as
\begin{eqnarray}
\label{eqs:R1}
R_1  &\equiv& \frac{\sigma(p_T^{\rm miss}<p_T^{\rm peak})}{\sigma(p_T^{\rm miss}>p_T^{\rm peak})}, \\
R_2  &\equiv& \frac{\sigma(p_T^{\rm miss}<p_T^{\rm cut})}{\sigma(p_T^{\rm miss}>p_T^{\rm cut})}.
\label{eqs:R2}
\end{eqnarray}
Here $p_T^{\rm peak}$ is defined theoretically as the $p_T$ value around which the distribution is the largest
(the experimental value of $p_T^{\rm peak}$ is discussed in the following).
It may take different values for the signal and background events.
And the default value of $p_T^{\rm cut}$ is chosen to be $150 ~{\rm GeV}$.
$R_1$ describes the profile of the peak region while $R_2$ incorporates the information on the tail region.
It is more convenient to adopt these two variables rather than the full distributions to understand the underlying processes.
Moreover, because of the ratios in $R_1$ and $R_2$, the coefficients of the operators are canceled.
As a consequence, $R_1$ and $R_2$ are functions of only the DM mass, providing a handle to the masses of invisible particles.
In particular, many effects that may change the leading-order prediction,
such as the factorization and renormalization scales,
PDF sets, parton shower, and higher order corrections are supposed to be canceled substantially in these observables as well.

We emphasize again that we are interested in how to distinguish the signal processes between Higgs invisible decay and DM associated production
after the discovery of the signal.
In this case, we can divide the total events observed experimentally to
the background (mainly $ZZ$ production) and signals (possible $HZ$ production or $SSZ$ production).
And therefore, the two observables, $R_1$ and $R_2$, can be measured separately for the background and signals.

\begin{figure}
\includegraphics[width=0.9\linewidth]{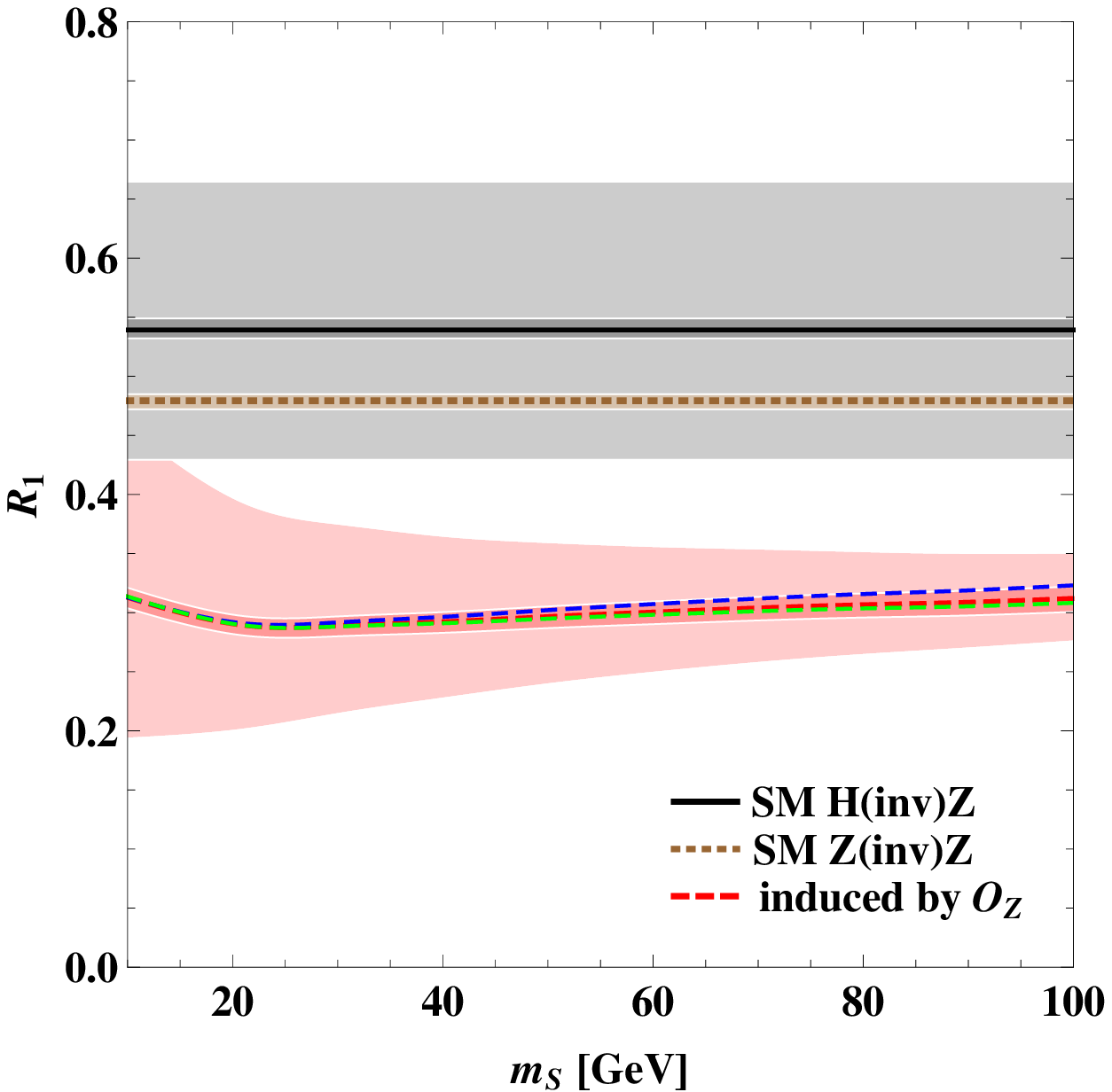}\\
\includegraphics[width=0.9\linewidth]{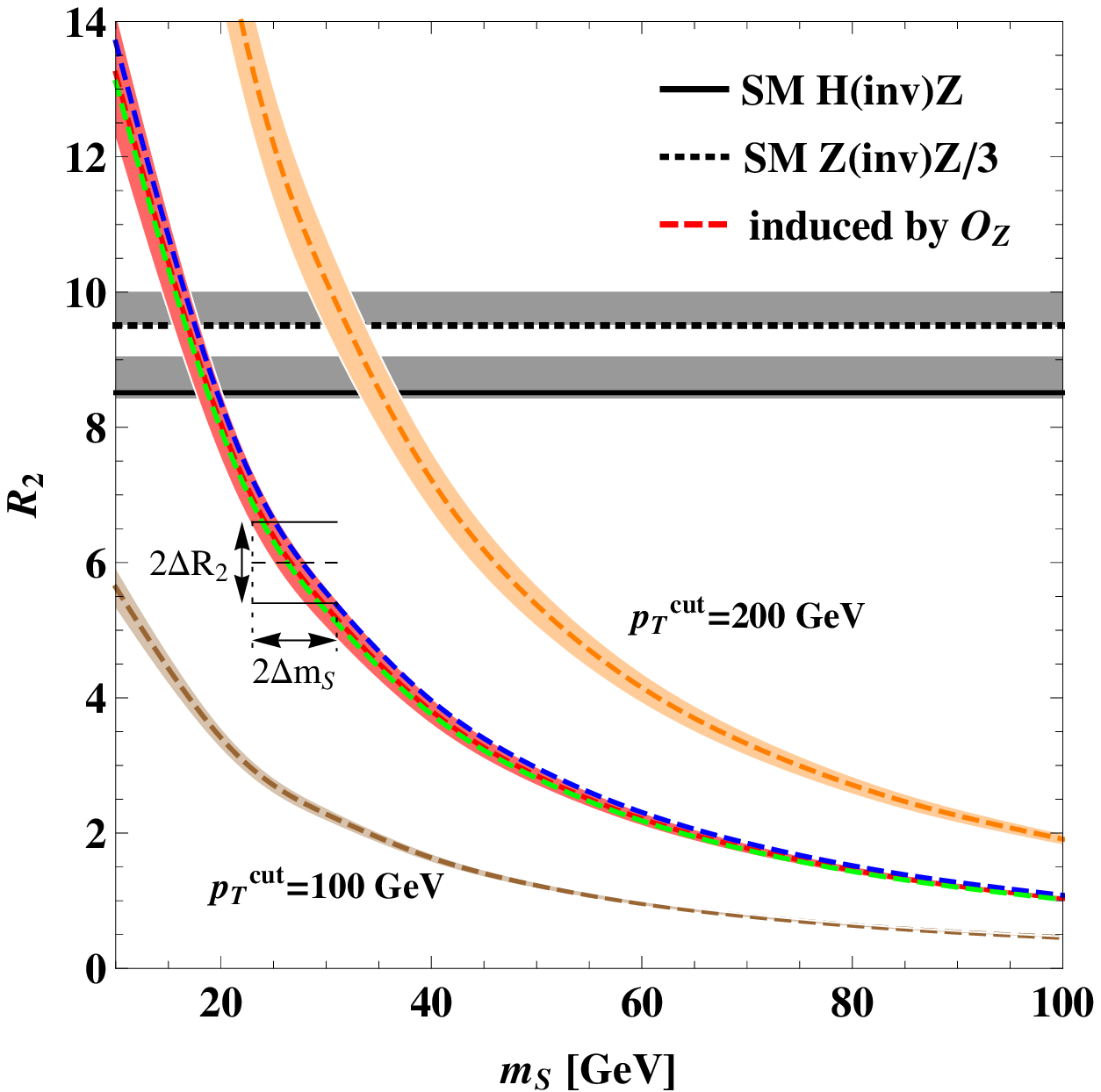}\\
  \caption{The dependence of  $R_{1}$ and $R_{2}$  on the DM mass.
  The narrow bands indicate the scale uncertainties by varying the default scale by a factor of 2.
  The wide bands in the upper plot are obtained by changing the $p_{T}^{\rm peak}$ by $\pm 5$ GeV.
  The blue and green dashed lines represent the results obtained by using CT10 and MSTW2008 PDF sets, respectively.
  The $R_2$ value for $Z({\rm invisible~decay})Z$  production has been divided by a factor of 3 in the bottom plot.
  In the example illustrating the estimation of $\Delta m_S$ in the bottom plot, $\Delta R_2$ is set to be $0.1 R_2$. }
  \label{fig:R_SSZB}
\end{figure}

The theoretical prediction for the dependence of $R_{1}$ and $R_{2}$ on the DM mass is shown in Fig.\ref{fig:R_SSZB}.
In the plot of $R_1$, the curves are obtained by using Eq. (\ref{eqs:R1}) with the corresponding $p_T^{\rm peak}$
given in Eqs.(\ref{eqs:pTpeak1}$\sim$\ref{eqs:pTpeak3}).
It can be seen that $R_1$ is insensitive to the DM mass for the processes induced by $\mathcal{O}_Z$
and takes discrepant values between the SM processes and processes induced by $\mathcal{O}_Z$.
On the other hand, the experiment would give $p_T$ distributions in bins for both the background and signal,
from which we can obtain a rough estimate of $p_T^{\rm peak}$.
But because of the limited signal events, the bin is perhaps too wide (for example 10 GeV) to precisely determine $p_T^{\rm peak}$.
In this case, we choose $p_T^{\rm peak}$ to be the left (right) edge of the maximum bin if the left (right)
neighbor of the maximum bin is larger than the right (left) one.
Then we count the left and right bins, obtaining the experimental value for $R_1$.
Given that the definitions of $p_T^{\rm peak}$ are different from the theory and experiment sides,
it is possible that they are not equal to each other.
This uncertainty would dilute the precision in determining the value of DM mass.
From Eq.(\ref{eqs:pTpeak1}), $\Delta m_S \approx 1.5 \Delta p_{T,Z}^{\rm peak}$.
Thus, an uncertainty of 5 GeV in $p_{T,Z}^{\rm peak}$ would result in an uncertainty of about 7.5 GeV in $m_S$.
So we will not use $p_{T,Z}^{\rm peak}$ to probe the value of $m_S$.
However, we can still use $R_1$ to separate the $HZ$ production with Higgs boson invisible decay and
$Z$ boson with DM associated production processes.
In Fig.\ref{fig:R_SSZB}, we show the uncertainty of $R_1$ by changing the $p_{T}^{\rm peak}$ by $\pm 5$ GeV.
It can be seen that the values of $R_1$ for the DM associated production and Higgs invisible decay processes do not overlap,
except for a very small region.

\begin{figure}
\includegraphics[width=0.9\linewidth]{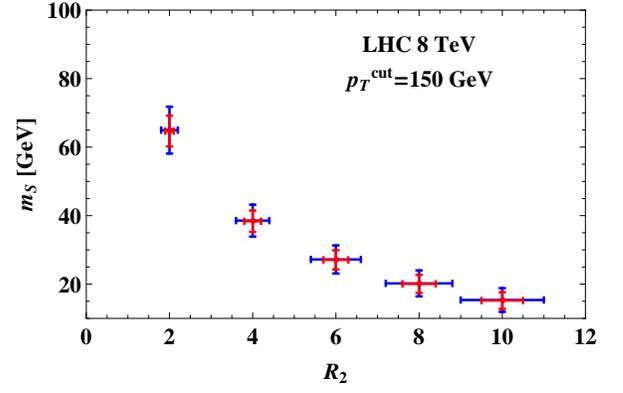}\\
  \caption{The determined $m_S$ as a function of  $R_{2}$ in the case of $p_T^{\rm cut}=150$ GeV.
  The error of $R_2$ is set to be $0.1 R_2$ and $0.05 R_2$ for the large and small error bars respectively,
  while the error of $m_S$ is derived as illustrated in Fig.\ref{fig:R_SSZB}. }
  \label{fig:error150}
\end{figure}

$R_2$ is very sensitive to the DM mass for the process induced by $\mathcal{O}_Z$, especially in the DM mass range $m_S<50$ GeV.
If the experimentally measured value of $R_2$ intersects with the curve corresponding to the process induced by $\mathcal{O}_Z$,
we can determine the mass of DM.
The uncertainties arising from the variation of scales and PDF sets are also calculated,  explicitly shown in Fig.\ref{fig:R_SSZB},
which turn out to be small, as expected.
After taking into account the scale uncertainties, which are much larger than those from the PDF sets,
the accuracy of the determined DM mass is estimated to be $\Delta m_S= 3.4\sim 6.8 $ GeV  as $R_2$ changes from 10 to 2
(corresponding to $m_S$ from 15.4 to 65.0 GeV); see Fig.\ref{fig:error150}.
This estimation is obtained without considering the effects of parton shower, hadronization and detector simulation
but under the assumption that the measured $R_2$ is accompanied with an uncertainty of $0.1 R_2$.
As the integrated luminosity is increased, the uncertainty of $R_2$ would be reduced.
We also show the situation in which the uncertainty of $R_2$ is $0.05 R_2$ in Fig.\ref{fig:error150}.
Then, the accuracy would be $\Delta m_S= 2.4\sim 4.5 $ GeV   as $R_2$ changes from 10 to 2.

\begin{figure}
\includegraphics[width=0.9\linewidth]{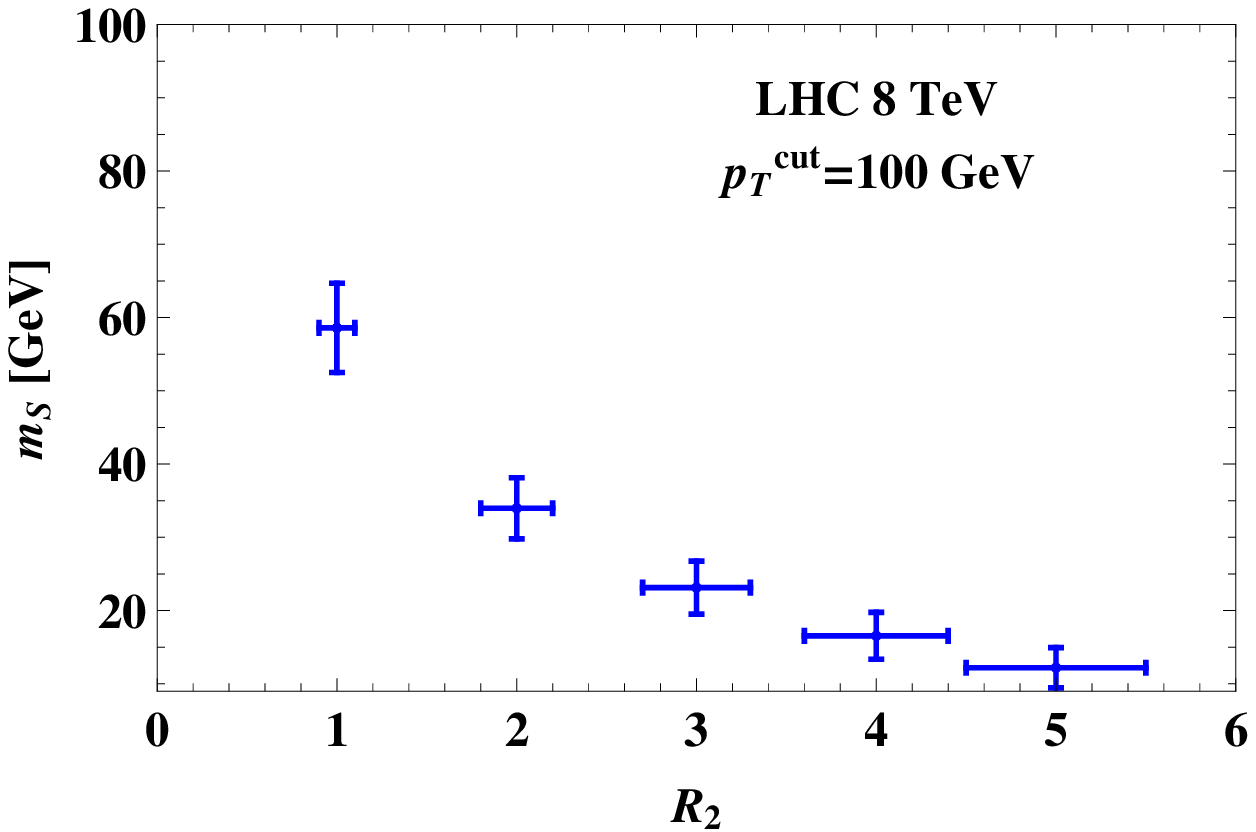}\\
\includegraphics[width=0.9\linewidth]{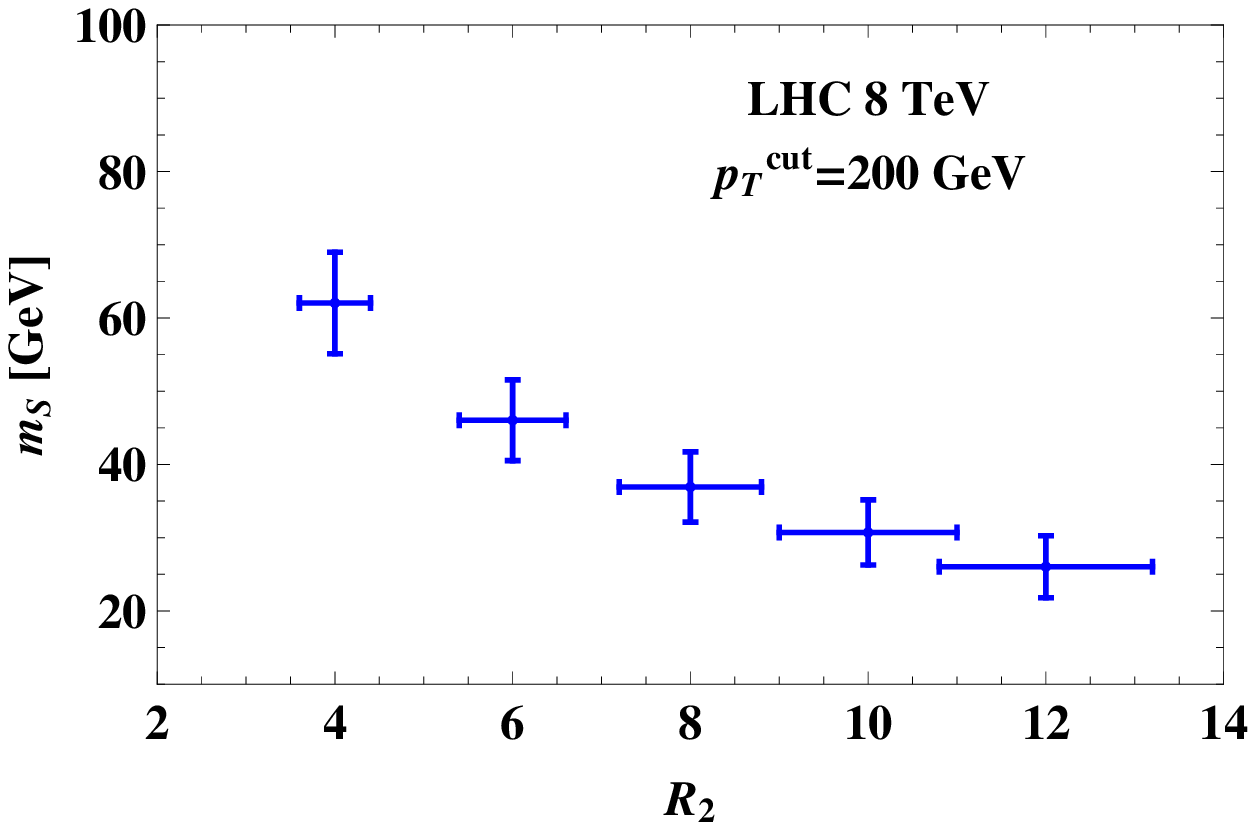}\\
  \caption{The determined $m_S$ as a function of  $R_{2}$ in the case of $p_T^{\rm cut}=100$ and 200 GeV.
  The error of $R_2$ is set to be $0.1 R_2$  while the error of $m_S$ is derived as illustrated in Fig.\ref{fig:R_SSZB}. }
  \label{fig:error100}
\end{figure}

Then we discuss the impact from the choice of $p_T^{\rm cut}$ in the definition of $R_2$ on the determination of the DM mass.
We have chosen $p_T^{\rm cut}=150$ GeV in the above numerical results.
But it is possible to choose a different value of $p_T^{\rm cut}$ as long as it is in the tail region (much greater than  $p_T^{\rm peak}$).
The two cases of $p_T^{\rm cut}=100$ and 200 GeV are also shown in Figs.\ref{fig:R_SSZB} and \ref{fig:error100}.
We see that for larger $p_T^{\rm cut}$, the scale uncertainty of $R_2$ at fixed $m_S$ is larger.
But $R_2$ drops faster as  $m_S$ increase at the same time.
The net effect results in a similar accuracy in estimating $m_S$.
We also notice that too large $p_T^{\rm cut}$ would induce a large statistical uncertainty,
and too small $p_T^{\rm cut}$ would reduce the sensitivity to $m_S$
\footnote{This sensitivity can be understood as $dR_2/dm_S$.}.
As a consequence, $p_T^{\rm cut}=150$ GeV is a good choice.

\section{conclusion}

To know the total width of the recently discovered Higgs boson particle, it is important to measure the invisible decay width of the Higgs boson.
However, the signal for this measurement at the LHC,
i.e., a charged lepton pair and missing energy in the final state, cannot be definitely understood as the product of
the intermediate produced $Z$ and $H$ bosons due to the possible interaction between DM and $Z$ boson or quarks,
which can be described by representative effective operators.
First, we consider the relic abundance, the LUX direct detection experiment and
the result of searching for Higgs boson invisible decay at the LEP and LHC in order to find the allowed parameter region space for the effective operators.
We discover that the interaction between DM and quarks is stringently constrained.
Then we investigate the transverse momentum distribution of the missing energy and propose two observables that can be used to distinguish
the different underlying processes.
Moreover, with these two observables, we may be able to determine the mass of invisible particles.

In this paper, we only consider the Higgs boson invisibly decaying into scalar DM,
inspired by the possible renormalizable extension of the SM.
And we also assume the $Z$ boson and quarks also interact with this kind of DM.
Given that the DM takes more parts of the energy in the Universe than ordinary matter, it is likely
that there are many kinds of DM
and the kind coupling with the $Z$ boson and quarks differs from that with the Higgs boson.
For example, the $Z$ boson and quarks connect with fermionic DM, while the
Higgs boson decays into scalar DM.
We will explore these scenarios in the future.

\section{acknowledgement}

This work was supported by
the Cluster of Excellence
{\it Precision Physics, Fundamental Interactions and Structure of Matter} (Grant No. PRISMA-EXC 1098).

\bibliography{Mass_of_inv}{}
\bibliographystyle{unsrt}

\end{document}